\documentclass[a4paper,11pt]{article}
\usepackage{pos}
\usepackage{amsmath}
\usepackage{graphicx}
\usepackage{microtype}
\usepackage{float}
\def\xfitter{\texttt{xFitter}}

\title{Resummation-scale effects in DIS PDF fits with \xfitter}
\ShortTitle{Resummation-scale effects in DIS fits}

\author*[a,b]{Francesco Giuli}
\affiliation[a]{Universit\`a degli Studi Link,  Via del Casale di S. Pio V, 44, 00165\\
Rome, Italy}

\affiliation[b]{INFN Sezione di Roma Tor Vergata, Via della Ricerca Scientifica, 1, 00133,\\
Rome, Italy}

\abstract{Renormalisation-group evolution enters PDF determinations through the running coupling and the splitting functions. At fixed perturbative order, the solution of the evolution equations is not unique beyond the nominal accuracy, and this residual freedom can be probed through resummation-scale variations. This contribution summarises an NNLO study of this effect in inclusive HERA DIS fits performed with the \xfitter{} fitting framework. The induced shifts are sizeable in a low-scale baseline fit, especially for the gluon and sea PDFs, and can affect LHC and FCC top-quark pair predictions at the few-percent level. Raising the PDF starting scale and applying a higher DIS virtuality cut considerably reduces the effect, a conclusion also supported by treating the resummation scale as a free parameter constrained directly by the data.}

\FullConference{The 33rd International Workshop on Deep Inelastic Scattering and Related Subjects (DIS2026)\\
4--8 May 2026\\
Bologna, Italy}

\begin{document}
\maketitle

\section{Motivation}

Modern LHC analyses rely on PDFs whose uncertainties are small enough not to obscure precision tests of QCD and electroweak physics. In phenomenological applications, missing higher orders are usually estimated by changing the renormalisation and factorisation scales that appear in the partonic hard cross sections. PDF evolution introduces another perturbative ingredient: the solution of renormalisation-group equations (RGEs) for the running coupling and for the parton densities. The ambiguity associated with this solution is formally of higher order, but it can still be numerically relevant when data at comparatively low scales are included in a fit.

A practical way to quantify this ambiguity was proposed in Refs.~\cite{Bertone:2022sso,Bertone:2024snr}. It introduces resummation scales into the RGEs, in analogy with resummed calculations for threshold or transverse-momentum observables. The scale variation then probes terms that are absent at a fixed perturbative order. Here this idea is applied to a PDF determination from inclusive HERA deep-inelastic scattering (DIS) data~\cite{H1:2015ubc}, using the public \xfitter{} framework~\cite{xFitter:2022zjb,Alekhin:2014irh}. The aim is to assess how much the PDFs, and predictions derived from them, change when the RGE solution is varied, and whether this uncertainty can be constrained or mitigated by the choice of fit setup.

\section{RGE variation}

For a generic renormalised quantity $R$, the scale evolution can be represented as
\begin{equation}
  \frac{d\ln R(\mu)}{d\ln\mu}=\Gamma\!\left(\alpha_s(\mu)\right),
  \qquad
  \Gamma(\alpha_s)=\sum_{n=0}^{\infty}\Gamma_n\alpha_s^{n+1} .
  \label{eq:rge}
\end{equation}
When Eq.~\eqref{eq:rge} is truncated, equivalent all-order expressions no longer give exactly the same numerical result. Such differences are higher order with respect to the adopted accuracy and therefore provide a natural estimate of missing contributions in the evolution.

The prescription used in the study amounts to evaluating the evolution after the replacement $\mu \to \xi\mu$, where $\xi$ is a dimensionless resummation-scale factor. The envelope from $\xi=0.5$ and $\xi=2$ is taken as the default uncertainty estimate. A more extreme value, $\xi=4$, is used only to illustrate the behaviour of the fit quality in the scan of the starting scale. As an alternative, $\xi$ can also be left free in the fit and constrained directly by the data (Sec.~\ref{sec:freexi}). Unlike the conventional variations of $\mu_r$ and $\mu_f$ in a hard process, this procedure modifies the perturbative evolution of PDFs and of $\alpha_s$, and the resulting uncertainty accumulates over the whole evolution span rather than being localised around the hard-scattering scale.

\section{DIS fit configuration}
\begin{figure}[t]
\centering
\includegraphics[width=0.45\linewidth]{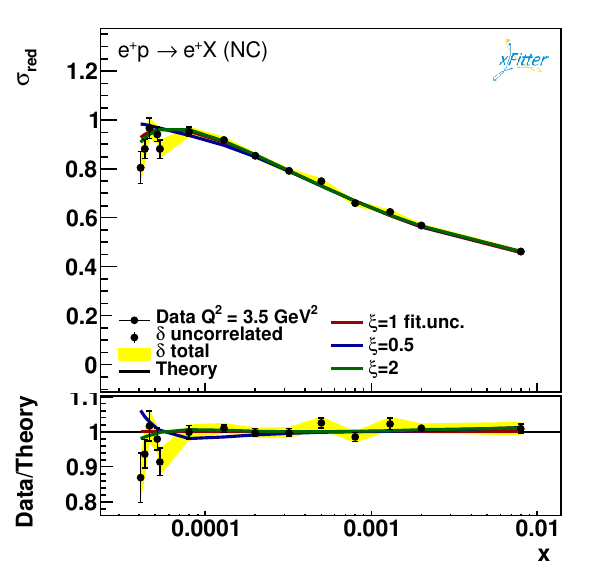}
\includegraphics[width=0.45\linewidth]{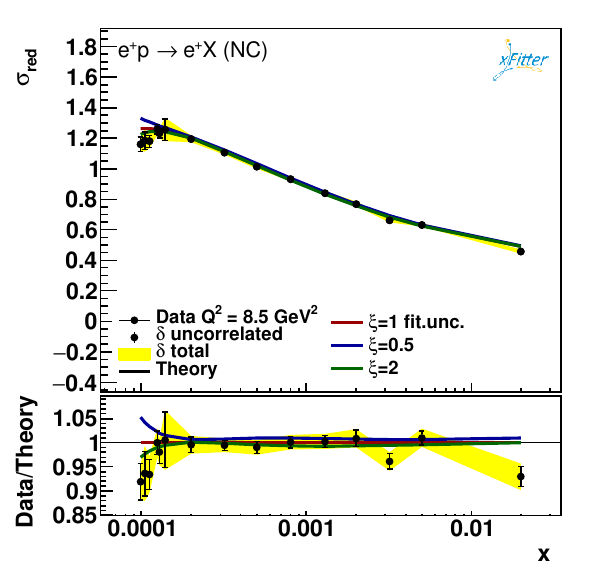}
\caption{Data-to-theory comparison for a representative set of HERA DIS NC reduced cross sections at $Q^2=3.5~\mathrm{GeV}^2$ (left) and $Q^2=8.5~\mathrm{GeV}^2$ (right). The bottom panels show the ratio theory/data for the nominal fit ($\xi=1$) and for the $\xi=0.5,2$ variations.}
\label{fig:datatheory}
\end{figure}
The fits use the combined inclusive HERA cross sections and follow the NNLO setup of Ref.~\cite{Bonvini:2019wxf}. Heavy flavours are treated in the FONLL-C scheme. The input parameters are fixed to $\alpha_s(M_Z)=0.118$, $m_c=1.46~\mathrm{GeV}$ and $m_b=4.5~\mathrm{GeV}$. At the input scale $Q_0$, five independent PDF combinations are parametrised: $xg$, $xu_v$, $xd_v$, $x\bar U$ and $x\bar D$. The functional forms can be found in Ref.~\cite{xFitter:2026}.
Sum rules fix the relevant normalisations. $B_{\bar U}$ and $B_{\bar D}$ are set equal, and the strange component is imposed with a constant suppression factor, $x\bar s=f_s x\bar D$, where $f_s=0.4$ and $A_{\bar U}=A_{\bar D}(1-f_s)$. After these constraints, 17 PDF parameters are fitted. Experimental PDF errors correspond to the $\Delta\chi^2=1$ criterion, obtained with the HESSE algorithm in \texttt{MINUIT}, and the $\chi^2$ minimisation itself is carried out with the \texttt{Ceres Solver} package.

Two fit variants are compared throughout. The reference configuration, labelled (1), starts the evolution at $Q_0=1.6~\mathrm{GeV}$ and includes data down to $Q^2_{\min}=3.5~\mathrm{GeV}^2$. A second configuration, labelled (2) and introduced to test the sensitivity to the low-scale region, uses $Q_0=3.2~\mathrm{GeV}$ together with $Q^2_{\min}=10~\mathrm{GeV}^2$. Figure~\ref{fig:datatheory} shows a representative comparison of the fitted theory to the HERA neutral-current reduced cross sections at low $Q^2$, together with the effect of the $\xi=0.5$ and $\xi=2$ variations; the impact on the theory curve is largest exactly in this low-$Q^2$, low-$x$ region.

\section{PDF response to the resummation scale}
\begin{figure}[t]
\centering
\includegraphics[width=0.45\linewidth]{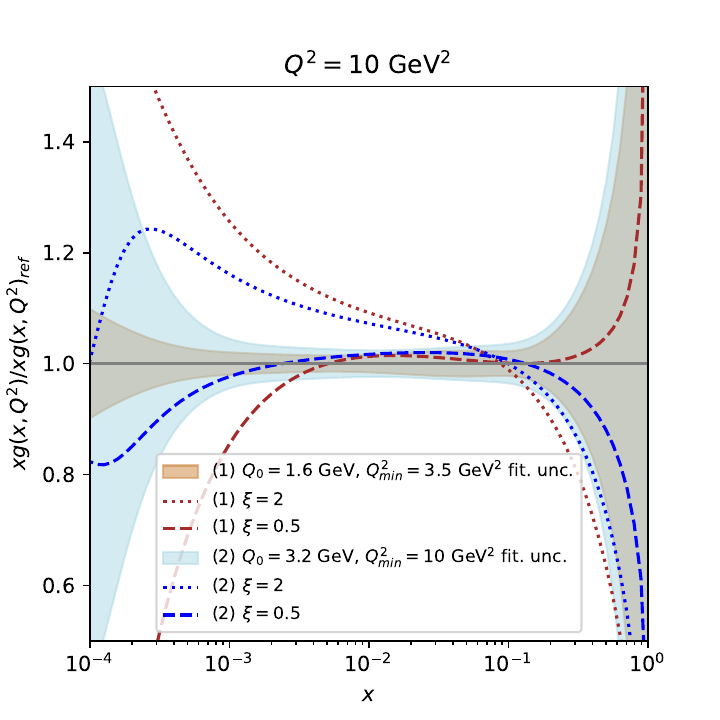}
\includegraphics[width=0.45\linewidth]{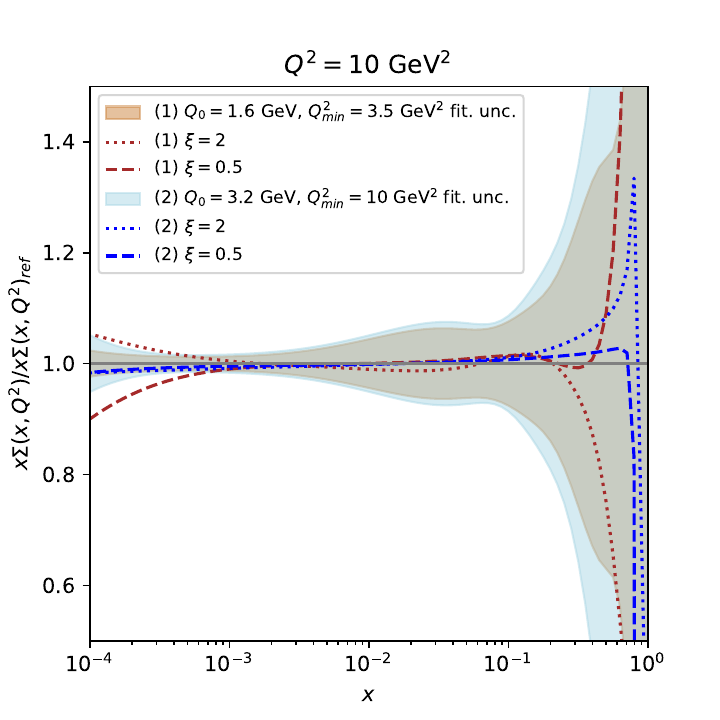}
\caption{Fitted gluon (left) and sea-quark (right) PDFs at $Q^2=10~\mathrm{GeV}^2$, obtained with $Q_0=1.6~\mathrm{GeV}$, $Q^2_{\min}=3.5~\mathrm{GeV}^2$ (1) and with $Q_0=3.2~\mathrm{GeV}$, $Q^2_{\min}=10~\mathrm{GeV}^2$ (2), for $\xi=0.5$ and $\xi=2$. Results are normalised to the corresponding central ($\xi=1$) fit, for which the fit uncertainty band is also shown.}
\label{fig:q2_10}
\end{figure}
\begin{figure}[t]
\centering
\includegraphics[width=0.45\linewidth]{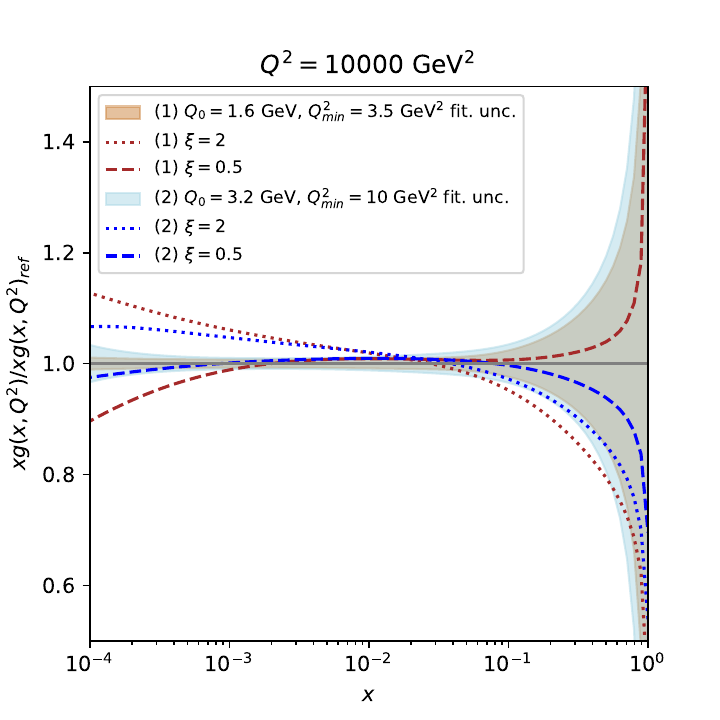}
\includegraphics[width=0.45\linewidth]{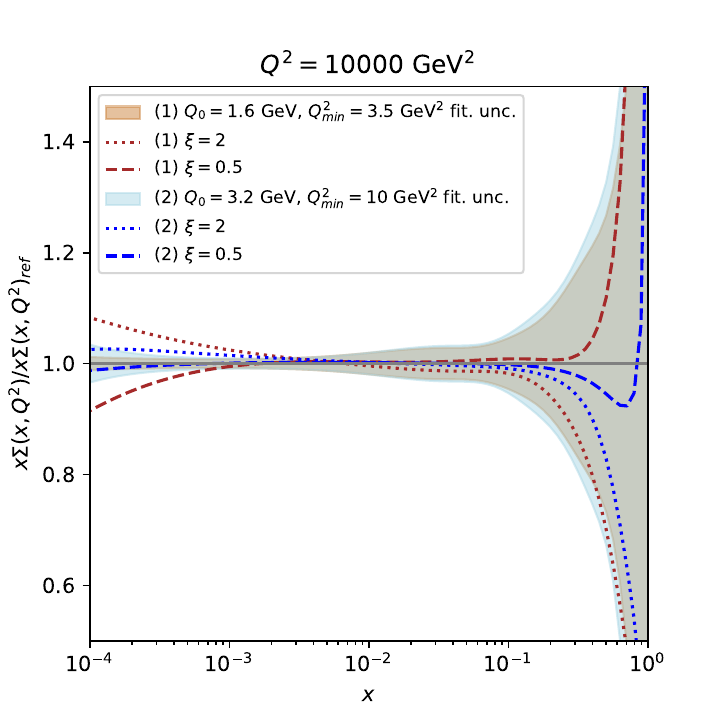}
\caption{As in Fig.~\ref{fig:q2_10}, but after evolution to $Q^2=10000~\mathrm{GeV}^2$.}
\label{fig:q2_10000}
\end{figure}
The PDF ratios for both fit setups at $Q^2=10~\mathrm{GeV}^2$ are displayed in Fig.~\ref{fig:q2_10}. The largest response is observed in the gluon and in the sea-quark ($\Sigma$) distributions of the reference setup (1): the changes induced by varying $\xi$ can be larger than the experimental fit uncertainty over a sizeable interval in $x$. Setup (2), with the higher starting scale and virtuality cut, shows a markedly smaller response over most of the $x$ range, confirming that the low-$Q^2$, low-$x$ region drives the effect.

Figure~\ref{fig:q2_10000} repeats the comparison after evolution to $Q^2=10000~\mathrm{GeV}^2$. The differences are generally milder in the well-constrained central-$x$ region, but the dependence on $\xi$ remains visible, especially at the edges of the $x$ range, and the hierarchy between setups (1) and (2) persists. Thus an uncertainty generated by the low-scale solution of the evolution equations is not confined to the fit scale: it can survive at the scales relevant for collider predictions.

\begin{figure}[t]
\centering
\includegraphics[width=0.45\linewidth]{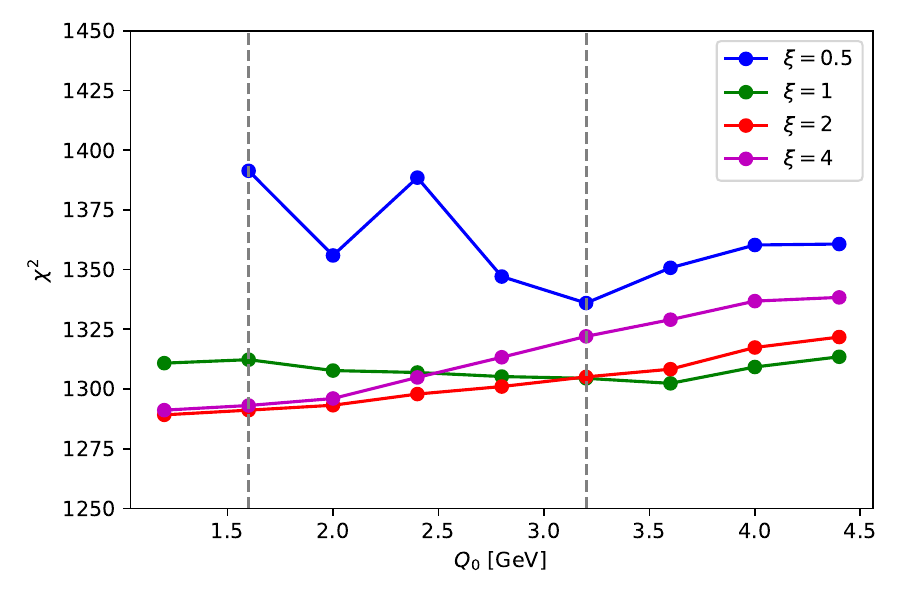}
\includegraphics[width=0.45\linewidth]{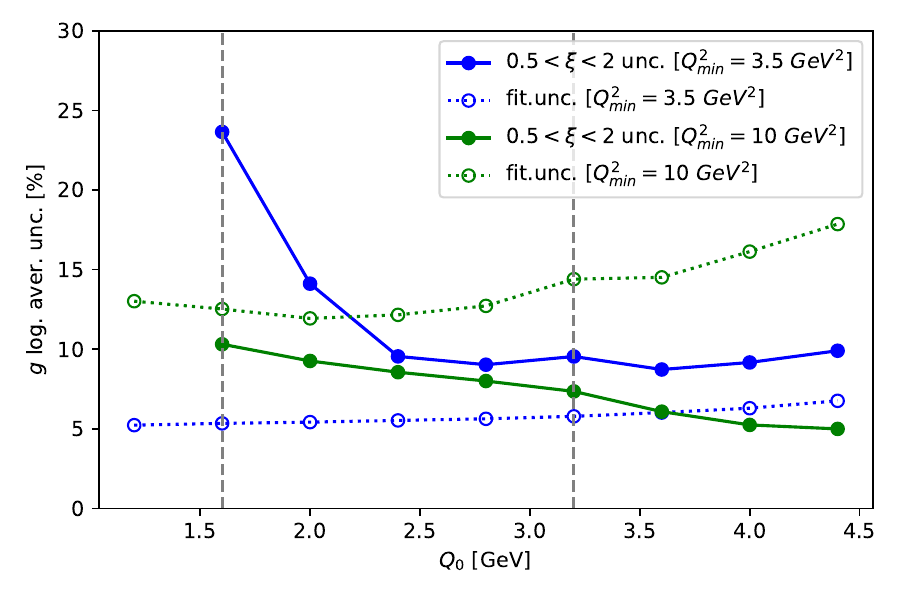}
\caption{Left: fit $\chi^2$ as a function of $Q_0$ for $\xi=0.5$, $1$, $2$ and $4$. Right: gluon-PDF fit uncertainty and $0.5<\xi<2$ resummation-scale envelope at $Q^2=10~\mathrm{GeV}^2$, as a function of $Q_0$, for the two virtuality cuts $Q^2_{\min}=3.5~\mathrm{GeV}^2$ and $Q^2_{\min}=10~\mathrm{GeV}^2$.}
\label{fig:q0scan}
\end{figure}

The dependence on the PDF starting scale is summarised in Fig.~\ref{fig:q0scan} (left). When $Q_0$ is close to the lowest perturbative scales, the fitted $\chi^2$ varies strongly with $\xi$, most notably for $\xi=0.5$. Around $Q_0\simeq3~\mathrm{GeV}$ the curves become closer, indicating a more stable evolution setup. The logarithmically weighted average uncertainty of the gluon PDF, shown as a function of $Q_0$ in Fig.~\ref{fig:q0scan} (right), confirms this trend directly: the $0.5<\xi<2$ resummation-scale envelope shrinks steadily with increasing $Q_0$, while the purely experimental fit uncertainty grows as the low-$Q^2$ data most sensitive to the gluon are progressively excluded. This trade-off between theoretical and experimental uncertainties motivates the choice $Q_0=3.2~\mathrm{GeV}$, $Q^2_{\min}=10~\mathrm{GeV}^2$ as the alternative fit setup, for which the resummation-scale uncertainty is substantially reduced (setup (2) in Fig.~\ref{fig:q2_10}), at the price of a somewhat larger fit uncertainty; in a global fit including high-energy collider data, the latter would be compensated by additional constraints on the gluon~\cite{Alekhin:2026det}.

\section{Resummation scale as a free fit parameter}
\label{sec:freexi}
An alternative way to account for the resummation-scale uncertainty is to treat $\xi$ as a free parameter in the PDF fit, allowing it to be constrained directly by the data, similarly in spirit to covariance-matrix approaches for theoretical uncertainties. The fitted values are $\xi=2.1\pm0.3$ for the reference setup (1) and $\xi=0.7\pm0.2$ for the alternative setup (2); in both cases the data prefer a value away from $\xi=1$, but the allowed range is considerably narrower than the conventional $0.5<\xi<2$ variation.

\begin{figure}[t]
\centering
\includegraphics[width=0.45\linewidth]{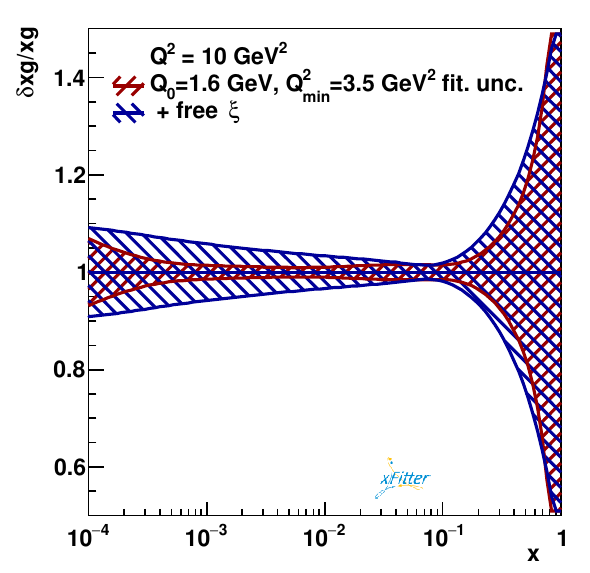}
\includegraphics[width=0.45\linewidth]{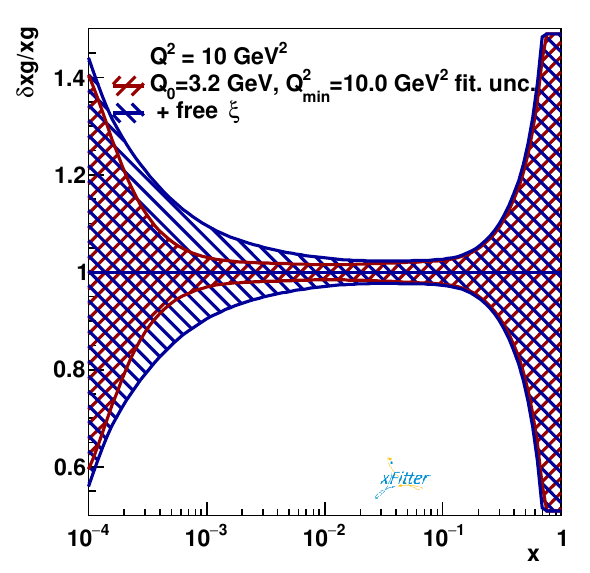}
\caption{Gluon PDF fit uncertainties at $Q^2=10~\mathrm{GeV}^2$, with and without a free $\xi$ parameter, for setup (1), $Q_0=1.6~\mathrm{GeV}$, $Q^2_{\min}=3.5~\mathrm{GeV}^2$ (left), and setup (2), $Q_0=3.2~\mathrm{GeV}$, $Q^2_{\min}=10~\mathrm{GeV}^2$ (right).}
\label{fig:freexi}
\end{figure}

Figure~\ref{fig:freexi} compares the resulting gluon PDF uncertainty, propagated from the fitted $\xi$, with the conventional fit uncertainty obtained by varying only the PDF shape parameters at fixed $\xi=1$. Letting $\xi$ float increases the PDF uncertainty noticeably, most prominently for the gluon and sea-quark distributions, though by a smaller amount than the full $0.5<\xi<2$ envelope of Fig.~\ref{fig:q2_10}. The same qualitative pattern as before is recovered: the additional uncertainty from the free $\xi$ parameter is markedly smaller for setup (2) than for setup (1), confirming that raising the starting scale and the virtuality cut stabilises the fit against the resummation-scale ambiguity irrespective of how the latter is estimated.

\begin{figure}[t]
\centering
\includegraphics[width=0.52\linewidth]{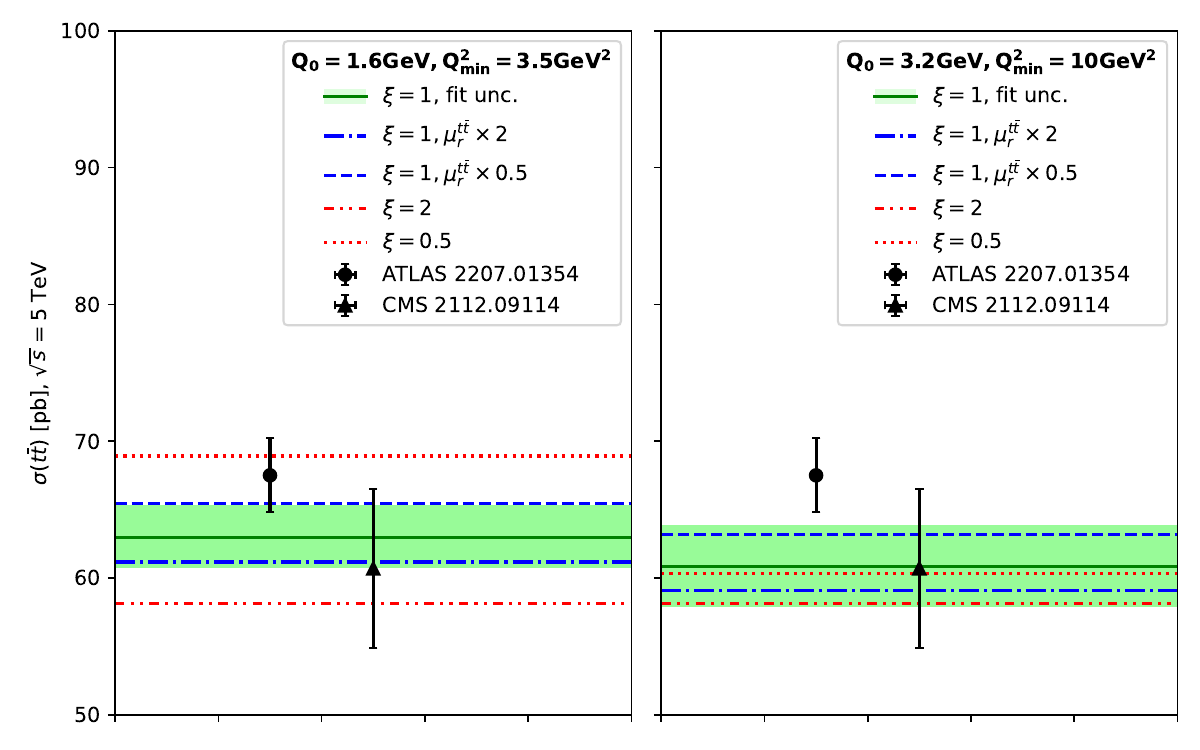}\\[-6pt]
\includegraphics[width=0.52\linewidth]{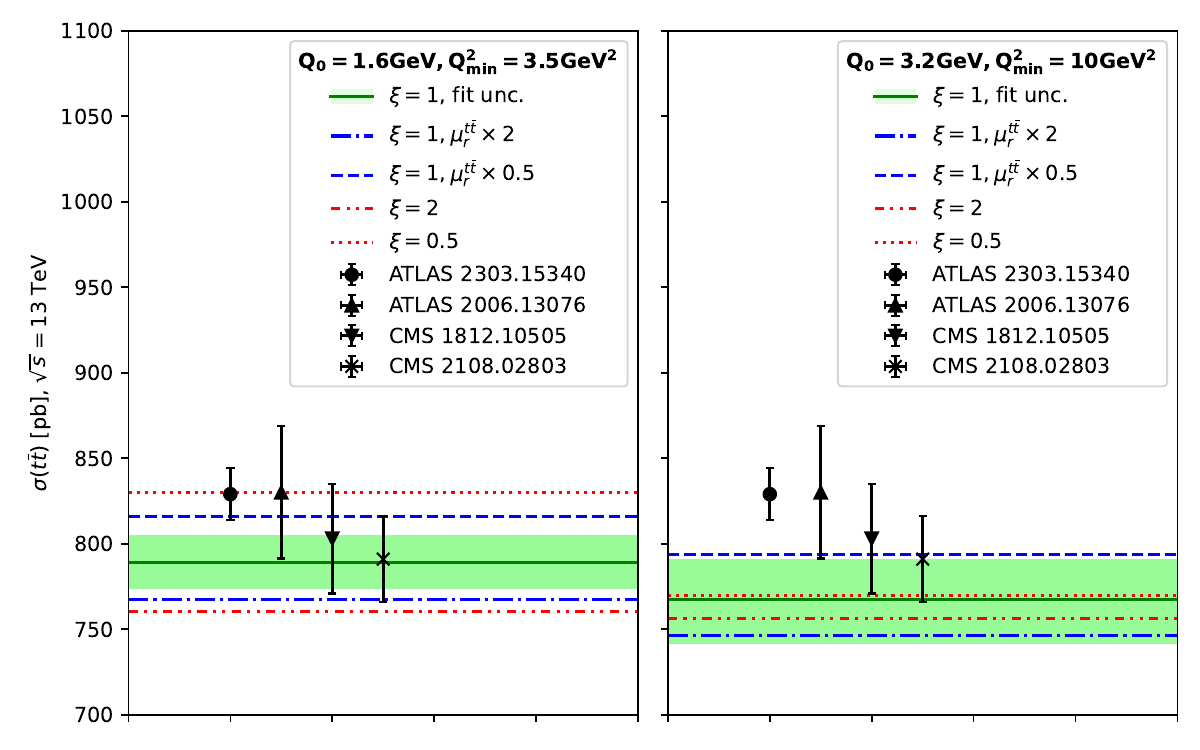}\\[-6pt]
\includegraphics[width=0.52\linewidth]{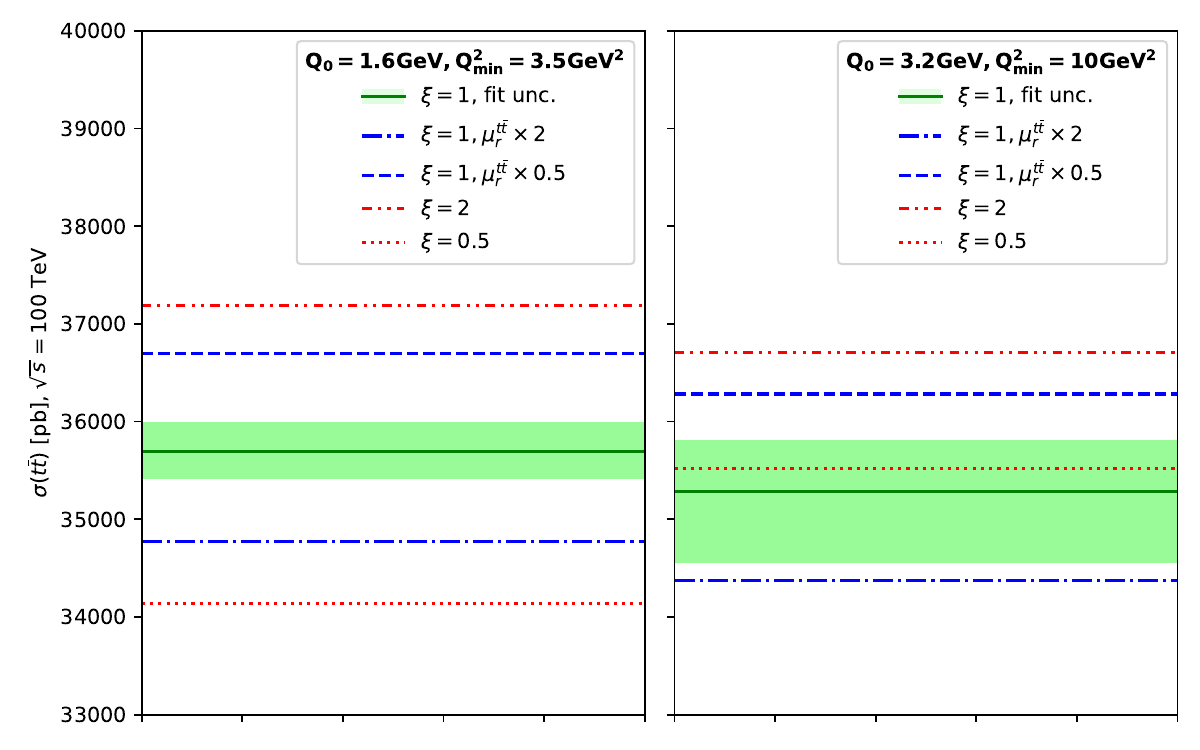}
\caption{Inclusive $t\bar t$ cross sections at $\sqrt{s}=5~\mathrm{TeV}$ (top), $13~\mathrm{TeV}$ (middle) and, at the FCC, $100~\mathrm{TeV}$ (bottom), computed with the PDFs from setup (1) (left column) and setup (2) (right column). The bands show the PDF fit uncertainty, the $\mu_r^{t\bar t}$ scale variation, and the $\xi=0.5,2$ resummation-scale variation; ATLAS and CMS data are overlaid where available.}
\label{fig:tt}
\end{figure}

\section{Effect on LHC and FCC observables}
The modified PDFs were propagated to the inclusive top-quark pair production cross section at $\sqrt{s}=5$, $13$ and $100~\mathrm{TeV}$, computed at NNLO QCD with \texttt{Hathor} as interfaced in \xfitter. The predictions are compared with representative ATLAS~\cite{ATLAS:2023gsl,ATLAS:2020aln} and CMS~\cite{CMS:2018fks,CMS:2021vhb} measurements in Fig.~\ref{fig:tt}, together with the conventional $\mu_r^{t\bar t}$ scale-variation band. In the reference setup (1), changing $\xi$ shifts the predicted cross section by roughly five percent, comparable to standard top-production scale variations and to present experimental uncertainties. Once the low-scale region is removed in setup (2), the corresponding shift is reduced to approximately one percent, while the benefits of threshold resummation are preserved.

The reduction is more pronounced at $\sqrt{s}=5~\mathrm{TeV}$ than at $\sqrt{s}=13~\mathrm{TeV}$: at the lower collider energy, $t\bar t$ production probes the gluon at larger $x$, a region constrained less directly by HERA data and more by the PDF extrapolation, which in turn depends on the momentum sum rule and hence on the small-$x$ shape most strongly affected by the resummation-scale variation. At the FCC energy of $100~\mathrm{TeV}$, where no data are yet available, the resummation-scale band remains a visible and non-negligible component of the total predicted uncertainty in setup (1), underlining the relevance of this source of uncertainty for future precision programmes.

\section{Summary}
Resummation-scale variations provide a way to assign an uncertainty to the fixed-order solution of RGEs used in PDF evolution. Applied to NNLO fits of inclusive HERA DIS data with \xfitter{}, this uncertainty is non-negligible in a conventional low-scale setup: it mainly affects the gluon and sea-quark PDFs, remains relevant after evolution to high scales, and produces a visible impact -- of order five percent -- on inclusive $t\bar t$ predictions at the LHC, with a comparable effect projected for the FCC. Treating the resummation scale as a free fit parameter gives a data-driven, and somewhat narrower, estimate of the same effect, without changing the qualitative conclusions. In both approaches, increasing the PDF starting scale and imposing a higher lower cut on $Q^2$ greatly improves the stability of the PDFs and of the derived collider observables, reducing the $t\bar t$ shift to about one percent while preserving the benefits of threshold resummation. These results motivate including RGE-solution uncertainties among the theory systematics considered in precision PDF determinations, particularly for future collider programmes.


\begin{thebibliography}{99}\small
\bibitem{Bertone:2022sso}
V.~Bertone, G.~Bozzi and F.~Hautmann,
Phys. Rev. D \textbf{105} (2022) 096003.

\bibitem{Bertone:2024snr}
V.~Bertone, G.~Bozzi and F.~Hautmann,
Phys. Rev. D \textbf{111} (2025) 074005.

\bibitem{H1:2015ubc}
H.~Abramowicz et al. [H1 and ZEUS Collaborations],
Eur. Phys. J. C \textbf{75} (2015) 580.

\bibitem{xFitter:2022zjb}
H.~Abdolmaleki et al. [\xfitter{} Developers' Team],
arXiv:2206.12465 [hep-ph].

\bibitem{Alekhin:2014irh}
S.~Alekhin et al.,
Eur. Phys. J. C \textbf{75} (2015) 304.

\bibitem{Bonvini:2019wxf}
M.~Bonvini and F.~Giuli,
Eur. Phys. J. Plus \textbf{134} (2019) 531.

\bibitem{xFitter:2026}
H.~Abdolmaleki et al. [\xfitter{} Developers' Team],
arXiv:2609.35344 [hep-ph].

\bibitem{ATLAS:2023gsl}
G.~Aad et al. [ATLAS Collaboration], arXiv:2303.15340 [hep-ex].

\bibitem{ATLAS:2020aln}
G.~Aad et al. [ATLAS Collaboration], arXiv:2006.13076 [hep-ex].

\bibitem{CMS:2018fks}
A.~M.~Sirunyan et al. [CMS Collaboration], arXiv:1812.10505 [hep-ex].

\bibitem{CMS:2021vhb}
A.~Tumasyan et al. [CMS Collaboration], arXiv:2108.02803 [hep-ex].

\bibitem{Alekhin:2026det}
S.~Alekhin, M.V.~Garzelli, S.~Moch and O.~Zenaiev,
Eur. Phys. J. C \textbf{86} (2026) 161 [arXiv:2510.21435 [hep-ph]].
\end{thebibliography}
\end{document}